\documentclass[english,usenatbib]{mn2e} 
\usepackage{amsmath} 
\usepackage{color}
\usepackage{babel} 
\usepackage{array} 
\usepackage{refstyle}
\usepackage{booktabs} 
\usepackage{units} 
\usepackage{multirow}
\usepackage{graphicx} 
\usepackage[unicode=true,pdfusetitle,
bookmarks=true,bookmarksnumbered=false,bookmarksopen=false,breaklinks,
pdfborder={0 0 1},backref=section,colorlinks=false,draft]{hyperref}

\usepackage{ulem}
\newcommand{\ed}[1]{{#1}}

\makeatletter
\providecommand{\tabularnewline}{\\}

\usepackage{aas_macros} 
\usepackage{epstopdf}
\PrependGraphicsExtensions{.jpeg} 
\PrependGraphicsExtensions{.svg} 

\makeatother
\begin{document} 

\title[High redshift supermassive blackholes: accretion through cold flows]{High redshift supermassive blackholes:
accretion through cold flows}
\author[Yu Feng et al.]{Yu Feng$^1$\thanks{yfeng1@andrew.cmu.edu}, 
  Tiziana Di Matteo$^1$,
  Rupert Croft$^1$ and Nishikanta Khandai$^2$ \\
  $^1$ McWilliams Center for Cosmology, Dept. of Physics, 
Carnegie Mellon University, Pittsburgh, PA, 15213 USA \\
  $^2$ Department of Physics, Brookhaven National Laboratory, Upton, NY 11973, USA
} 

\maketitle

\begin{abstract} 
  We use zoom-in techniques to re-simulate three high-redshift ($z \ge 5.5$)
halos which host $10^9$ solar mass blackholes 
from the $\sim$ Gpc volume, MassiveBlack cosmological hydrodynamic simulation.
We examine a number of factors potentially affecting supermassive blackhole
growth at high redshift in cosmological simulations. These include numerical
resolution, feedback prescriptions and formulation of smoothed particle
hydrodynamics. We find that varying the size of the region over which feedback
energy is deposited directly, either for fixed number of neighbours or fixed
volume makes very little difference to the accretion history of blackholes.
Changing mass resolution by factors of up to 64 also does not change the
blackhole growth history significantly. We find that switching from the
density-entropy formulation to the pressure-entropy formulation of smoothed
particle hydrodynamics slightly increases the accretion rate onto blackholes. In
general numerical details appear to have small effects on the main fueling
mechanism for blackholes at these high redshifts.  We examine the fashion by
which this occurs, finding that the insensitivity to simulation technique seems
to be a hallmark of the cold flow feeding picture of these high-$z$ supermassive
blackholes.  We show that the gas that participates in critical accretion
phases, in these massive objects at $z > 6\sim7$ is in all cases colder, denser,
and forms more coherent streams than the average gas in the halo. This is also
mostly the case when the blackhole accretion is feedback regulated ($z < 6$),
however the distinction is less prominent.  For our resimulated halos, cold
flows appear to be a viable mechanism for forming the most massive blackholes in
the early universe, occurring naturally in $\Lambda$CDM models of structure
formation. Not requiring fine tuning of numerical parameters, they seem to be
physically inevitable in these objects.
\end{abstract}

\begin{keywords} supermassive blackholes -- numerical scheme -- cosmological simulation 
\end{keywords}
\section{Introduction}
Deep sky surveys have revealed populations of 
distant quasars at redshifts $z > 6$,
\citep{
2001AJ....122.2833F, 2003AJ....125.1649F, 2004AJ....128..515F,
2006AJ....131.1203F,
2009AJ....138..305J,2009ApJS..182..543A, 
2011Natur.474..616M, 
2012AJ....143..142M, 
2013ApJ...770...13W}. 
The mass of the central blackholes in high redshift quasars are estimated 
to be 
$M_\mathrm{BH} \sim 10^9 M_\odot$; 
the feasibility of growing such blackholes on
a timescale of less than 1 billion years poses tight constraints on
astrophysical mechanisms for doing this.
Two aspects of growing such super-massive blackholes are seeding and the
subsequent accretion of mass.
On the seeding side, seed masses of from $100 - 10^5 \,M_\odot$ 
at redshifts $z>10$ have been proposed by various authors. 
On the lower end, they could form from the remnants of PopIII stars, 
\citep{1999ApJ...527L...5B,
2000ApJ...540...39A,2001ApJ...548...19N, 2003ApJ...598...73Y,2005MNRAS.363..379G}, or for the most massive seeds 
from direct gravitational collapse \citep{2004MNRAS.354..292K,
2006MNRAS.370..289B, 2010Natur.466.1082M,
2011ApJ...742...13B,
2013ApJ...774..149C}. 

Given a seed, the blackholes must 
endure a sustained period of Eddington limited 
growth \citep{
2005ApJ...633..624V,
2006MNRAS.371.1813L,
2007ApJ...665..107P} to
grow to masses of $10^9 M_\odot$. An important factor in studying the
growth of these objects is understanding whether 
sufficiently strong gas inflows are present in high redshift halos 
and how self-regulation and feedback from 
supermassive 
blakcholes 
may 
affect 
them 
\citep{2005Natur.433..604D,
2007MNRAS.376.1547C,
2008AN....329..956J,
2009MNRAS.398...53B,
2010Natur.466.1082M,
2011MNRAS.412.1341D,
2012MNRAS.421.3443H,
2012MNRAS.420.2662D}.

Numerical simulation 
of the growth of
supermassive blackholes in a cosmological volume is a
challenging problem. This is especially the case for the first
quasars. 
Constraints from the quasar luminosity function 
\citep[e.g.][]{2001AJ....122.2833F}
imply that such quasars are
rare objects, with a density of $\sim 1-10\, \unit{Gpc}^{-3}$. As a
consequence, to simulate the first quasars, a large volume on the order of
$\unit{Gpc}^{3}$ with sufficient 
mass resolution to  resolve their host galaxies and the gas inflow within 
them is required.

To work around this difficulty, many authors have adopted a method of
``resimulation'', where high resolution, ``zoomed-in'' initial conditions are
generated selecting high redshift massive halos selected from large volume,
lower resolution simulations \citep[see, e.g.,][]{
  2013MNRAS.428.2885D,
  2013arXiv1303.6959A,
  2013arXiv1307.0856B,
  2013arXiv1307.5854C, 2011ApJ...736...66R,
  2011ApJ...741L..33B, 
  2010MNRAS.407.1529H,2009MNRAS.400..100S,
  2007ApJ...665..187L}.
In this commonly adopted method however the growth of the most massive
black holes is then typically assumed to be associated with the most 
dark matter halos. This may not be the case at all redshifts:
it is still controversial whether
halo mass is a good indicator of the blackhole properties of the rare first
quasars \citep[see,
e.g.,][]{2013MNRAS.436..315F,2013MNRAS.432.2869H,2009ApJ...695..809K}.

Another approach is to directly run a large volume simulation with
hydrodynamics and a model for supermassive blackhole formation and
growth \citep[see,
e.g.,][]{2013NatSR....E1738B,
2013MNRAS.431.2513W,
2010MNRAS.405L...1B,2009MNRAS.398...53B,
2009MNRAS.396..423B, 
  2008MNRAS.383.1210S,
2008ApJ...676...33D,
2008MNRAS.383..289P}.
The largest simulation in this approach is the MassiveBlack (hereafter MB)
simulation by \cite{2012ApJ...745L..29D}, designed to study the first quasars.
With a $0.75\,\unit{Gpc}$ box side length and $2\times3200^3$ particles, the
Smoothed Particle Hydrodynamics (hereafter SPH) simulation runs from uniform
cosmological initial conditions produced around ten supermassive blackholes
with $M_\mathrm{BH} \sim 10^9 M_\odot$ at $ z \sim 6$. The growth of these
blackholes was found to be mainly due to cold flows. Gas in the vicinity of
the blackholes was shown to originate from cold dense filaments that survive
well within the virial radius of the halo  \citep[see
also][]{2011ApJ...741L..33B,2009Natur.457..451D}. Even though the mass
resolution of MB is sufficient to resolve the host halos of the first quasars,
the spatial resolution was not sufficient to follow gas inflows below sub-Kpc
scales. Another drawback with large uniform simulations such as MB is that it
becomes prohibitively expensive to experiment with the numerical schemes for
hydrodynamics and feedback.

In this work we apply the resimulation method to a sample of three halos
hosting $10^9 M_\odot$ blackholes in the hydrodynamical MB simulation. Our
main goal is to study the growth of these extreme objects at higher resolution.
We will examine the effect of different numerical schemes on the accretion
history of blackholes, then proceed to investigate 
how gas arrives to and participates in the growth of supermassive blackholes.

The paper is organized as follows: in Section 2 we discuss the selection and
construction of the initial conditions; in Section 3, we study how
method of feedback energy deposition, resolution and
SPH formulation affect the accretion history of the supermassive blackholes;
in Section 4 we investigate in detail the picture of cold flow feeding.

\section{Initial Conditions}
\label{sec:ic}
\subsection{Selection of Halos}
\begin{figure*}
  \includegraphics[width=\textwidth]{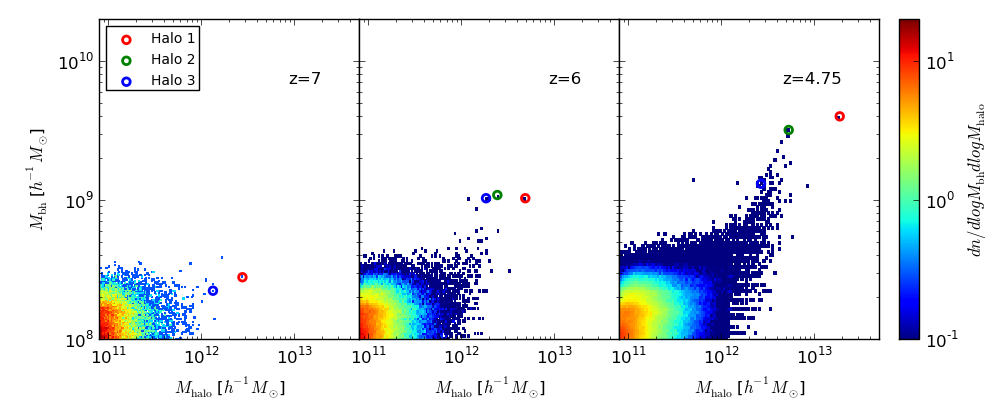}
  \caption{The mass of blackholes $M_\mathrm{bh}$ compared
to the total mass of their host dark
matter halos, $M_\mathrm{halo}$.
The color in the histogram represents the number density of
halos per logarithm of blackhole mass and per logarithm of halo mass.
The three halos we have targeted for resimulation
are marked with colored circles: red is halo 1, green is halo 2, and blue is
halo 3. The blackhole in halo 3 at $z=7$ is not shown in the figure because 
its mass is less than $10^8\unit{M_\odot}$.
}
      \label{fig:mbh-mhalo}
\end{figure*}
\begin{figure*}
  \includegraphics[width=2.2in]{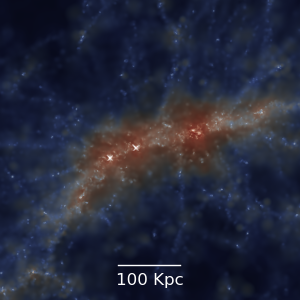} %
  \includegraphics[width=2.2in]{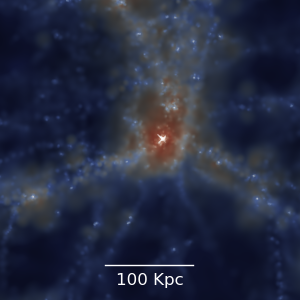} %
  \includegraphics[width=2.2in]{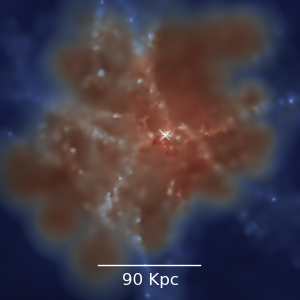}
  \caption{The environment of the halos targeted
for resimulation as seen in the MB simulation
at redshift $z=6$. 
From left to right: Halo 1, Halo
  2 and Halo 3. Shown is the projected gas density color coded by 
temperature (red $=10^8 \unit{K}$, blue $=10^4 \unit{K}$). 
The scale in each panel marks the virial radius (typically \~ 100 Kpc 
at $z=6.0$.}
  \label{fig:halo-env}
\end{figure*}

The MB simulation
can be used to examine the relationship between blackhole mass and 
halo mass for the entire population of hosting halos, and this is shown in 
Figure \ref{fig:mbh-mhalo} at three redshifts $z=7$, $z=6$, and $z=4.75$.

We can see that for the 
same host mass $M_\mathrm{HALO}$ of a few times
$10^{12} M_\odot$, the blackhole mass can vary
significantly (by greater than an order of magnitude). Similar scatter is
also reported by \cite{2013MNRAS.436..315F} from semi-analytic modelling of 
blackhole growth in dark matter simulation.
The scatter shows that picking target halos from a hydrodynamic simulation
is necessary to ensure that we are sampling the distribution of
supermassive blackholes that we want. 

We select three halos from the MB simulation based on their
blackhole mass at $z=6$.  We pick 3 of the most massive
blackholes (shown as circles in Figure \ref{fig:mbh-mhalo}). 
Their large scale gas envrionment is shown in figure \ref{fig:halo-env}.
The three halos evolve differently even though they have a similar mass of $\sim
10^{13}\,\mathrm{M_\odot}$ at $z=6$.  
The ranking of halos by blackhole mass can change significantly between
redshifts. For example, at $z=7$ the blackhole mass in halo 2 does not even show
up in Figure \ref{fig:mbh-mhalo} ($M_\mathrm{BH} < 10^8\,\mathrm{M_\odot}$), but
at $z=6.0$ it becomes one of the most massive blackholes ($ M_\mathrm{BH} \sim
10^9\,\mathrm{M_\odot}$).

We now describe the environment and blackholes in each halo in turn (see also
Figure \ref{fig:bh-mergers}):
\begin{enumerate}
    \item[Halo 1] 
      lies along the most prominent filament in the MB simulation (first panel
      in Figure \ref{fig:halo-env}. The halo eventually amalgamates with nearby
      halos into one ``super halo'' that is several times more massive than any
      other halos in the simulation. 
      The most massive progenitor blackhole grew rapidly at high redshift $z=9$ 
      via a continuous supply of cold gas flows through the filament. At low
      redshift the blackhole mass increased due to a major merger event. 
      Two nearby blackholes are also approaching the most massive blackhole,
      though they have not merged at $z=4.75$ (end of simulation).
    \item[Halo 2] is located at the confluence of 3 filaments (second panel in
Figure \ref{fig:halo-env}). From a general
      visualization of the MB simulation \citep{2011ApJS..197...18F} we can
      infer that this is one of the typical configurations for massive halo
      growth by accretion from filaments. The blackhole went though 
      a rapid growth phase between $z=7$ and $z=6$, likely due to an 
      incoming gas rich merger. 
    \item[Halo 3] is hosted by a
      quiescent environment.  The growth of the blackhole has been quenched
      at $z=7$ by feedback which has reduced the cold gas supply. As we can see
from the third panel in Figure \ref{fig:halo-env}, the halo
      is surrounded by a large region of hot gas.
\end{enumerate}

\subsection{Generating the Initial Conditions}
\begin{figure}
  \centering
  \includegraphics[width=0.8\columnwidth]{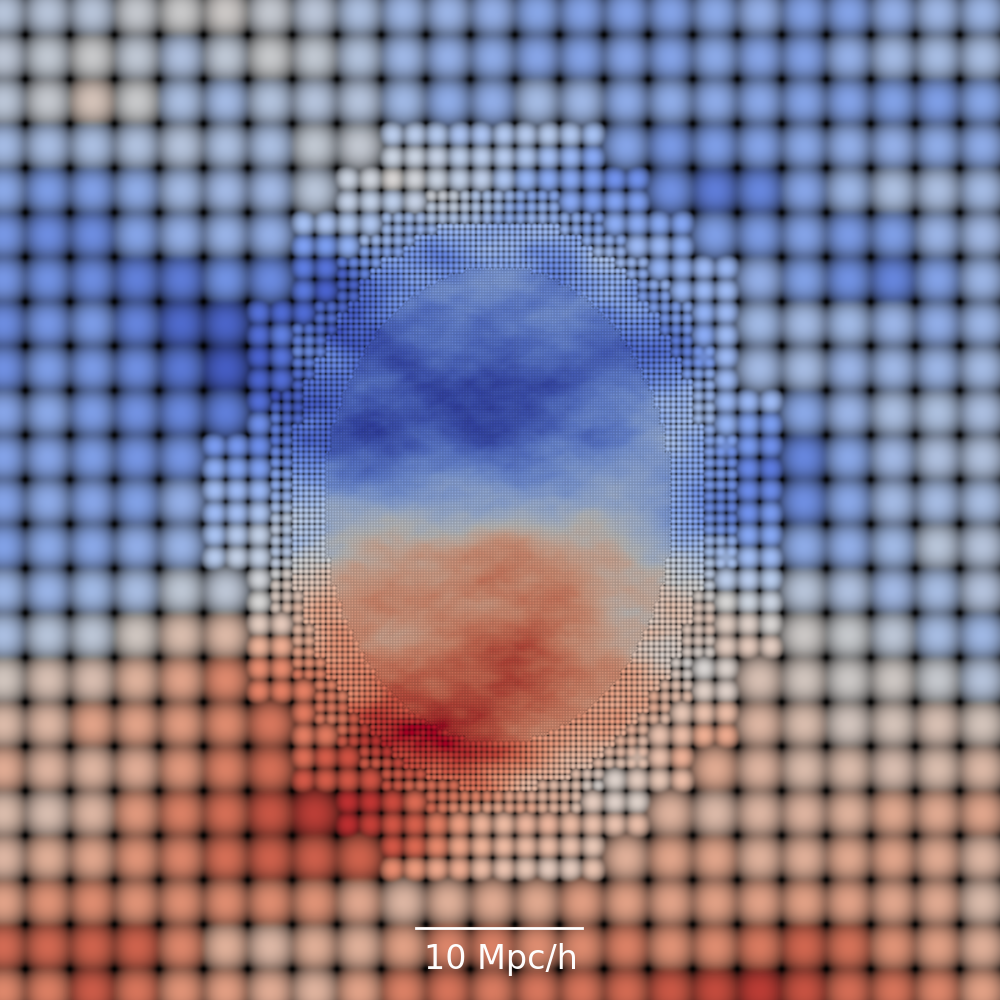}
  \caption{Illustration of gradual degrading of resolution of
   initial conditions with 2-d projection of a slice along $z$-direction. 
   Color represents the initial Zel'Dovich $x$-velocity; the points represent
particles: larger points correspond to lower resolution; in this particular set
up, there are 5 resolution levels. For illustration purpose, the particles in
the figure are not displaced by their initial displacement.  (The initial
conditions are.) The large scale fluctuation blends smoothly across the boundary
between zoom regions as seen for the $x$-velocity field. The total mass and
center of mass of particles are conserved by ensuring that whenever a lower
resolution particle is replaced by a higher resolution particle, it is fully
replaced by 8 particles (shown as four in the 2-d projection).
   }
  \label{fig:IC}
\end{figure}

The cosmological parameters used for the initial conditions are identical to that
used in MB (listed in Table \ref{tab:cosmology}). 
\begin{table}
  \centering 
  \begin{tabular}{ccccccc}
    \hline
    \hline
    $h$ & $\Omega_b$ & $\Omega_\Lambda$ & $\Omega_M$ & $\sigma_8$ & $z_\mathrm{init}$ & $z_\mathrm{end}$\\
  \hline
  0.72 & 0.044  &  0.74    & 0.26 & 0.8 & 159.0 & 4.75 \\
  \hline
  \hline
  \end{tabular}
  \caption{Cosmology Parameters}
  \label{tab:cosmology}
\end{table}

Our algorithm for generating the initial conditions uses
as a base the code from the initial condition generator
that was used for the MB simulation
(N-GenIC, author Volker Springel). This is  
in order to preserve the same randomly
sampled large scale Fourier modes 
used in MB. 
We modified N-GenIC to produce multiple levels of 
displacement at a selected region, and implemented a post-processing algorithm
that assembles the particles and their displacement at different resolution
levels while conserving the total mass and the center of mass of particles.

To define the high resolution zoom region we first
find the dark matter particles in the friends-of-friends \citep[FOF, see][]{
1985ApJ...292..371D} group
corresponding to the selected halo at redshift $z=6$. We then find the initial
positions of these particles, and put down a bounding sphere. We enlarge the
bounding sphere by a factor of 1.5, and the region inside this sphere is
our highest resolution zoom region. We populate this region 
with high resolution gas
and dark matter particles whose initial perturbation is the sum of the
interpolated (4th order spline) large scale mode perturbations in
the original MB initial conditions 
and additional small scale mode perturbations (with random
Fourier phases) drawn from smaller boxes.
Outside the spherical high resolution region, a series of 
spherical shells are populated with lower resolution matter particles
(collisionless), each shell
radius being a factor of 1.14 larger than that 
interior to it, until we have reached
a spatial resolution 8 times worse than that of MB. 
The rest of the cubical
volume is populated with particles of this mass. The setup is illustrated with a
2d-projection of the $x$ velocity field in Figure \ref{fig:IC}.

During the simulation, we measure the minimal distance from the nearest low
resolution particle to the central blackhole. The closest distance is
$1 h^{-1}\unit{Mpc}$ in co-moving units, which is greater than the
virial radius of the halo. This provides enough evidence that particles from the
low resolution regions do not contaminate the blackhole accretion and halo
growth.

\section{Simulations}
\label{section:sims}
Our fiducial SPH simulation code is the same as that used
to run the original MB model
\citep{2012ApJ...745L..29D}, {\small P-GADGET3}
\citep[see][for details of the hydrodynamics and gravity computation]
{2005MNRAS.364.1105S}.
The formulation of SPH used in MB is density-entropy SPH 
\citep{2002MNRAS.333..649S,
1977AJ.....82.1013L,1977MNRAS.181..375G}, 
with
a cubic spline
kernel, the blackhole accretion model from \cite{2005Natur.433..604D} and
the multiphase star-formation model from \cite{2003MNRAS.339..289S}. 
In order to assess the 
effects of choices for the numerical schemes, we 
incorporate additional features into the
code: \begin{itemize}
  \item a fixed feedback deposition volume in the blackhole accretion model;
  \item an alternative, pressure-entropy SPH formulation to alleviate the
  problem of unphysical SPH surface tension
    \citep{2013MNRAS.428.2840H, 2010MNRAS.405.1513R};
  \item a quintic smoothing kernel to alleviate the problem of pairing instability; 
    we also change the variable controlling the SPH resolution from
    $N_\mathrm{NGB}$ (number of nearest neighbours) to
    $\eta$ (ratio between SPH resolution and mean particle separation)
    recommended by \cite{2012JCoPh.231..759P} \citep[see
    also][]{2012MNRAS.425.1068D}.
\end{itemize}

In this work we present 18 simulation with various combinations of these models
and the resolutions. The matrix of the simulations and their abbreviated names
are listed in Table \ref{tab:simulations}. We note that our LDCA simulations use
the same model and resolution as MassiveBlack; and the HDCV simulations are the
simulations with the highest resolution (64 times finer mass resolution and 4
times finer spatial resolution than MB). 

\begin{table*}
  \centering 
  \begin{tabular}{ccccccc}
    \hline
    \hline
    Symbol & Halo & $\varepsilon_\mathrm{grav}$ $h^{-1}\unit{Kpc}$ & 
    $m_{0\mathrm{DM}}$ $\unit{M_\odot}$& SPH & Kernel & Feedback \\
    \hline\hline
    1HDCV &  1 & 1.5 &  $4.3\times10^6$ & DE & Cubic & Volume \\
    1MDCV &  1 & 3.0  & $3.5\times10^7$ & DE & Cubic & Volume \\
    1MDCA &  1 & 3.0  & $3.5\times10^7$ & DE & Cubic & Adaptive \\
    1MDQV &  1 & 3.0  & $3.5\times10^7$ & DE & Quintic & Volume \\
    1MPQV &  1 & 3.0  & $3.5\times10^7$ & PE & Quintic & Volume \\
    1LDCA$^\dagger$ &  1 & 5.5  & $2.8\times10^8$ & DE & Cubic & Adaptive \\
    \hline
    2HDCV &  2 & 1.5 &  $4.3\times10^6$ & DE & Cubic & Volume \\
    2MDCV &  2 & 3.0  & $3.5\times10^7$ & DE & Cubic & Volume \\
    2MDCA &  2 & 3.0  & $3.5\times10^7$ & DE & Cubic & Adaptive \\
    2MDQV &  2 & 3.0  & $3.5\times10^7$ & DE & Quintic & Volume \\
    2MPQV &  2 & 3.0  & $3.5\times10^7$ & PE & Quintic & Volume \\
    2LDCA$^\dagger$  &  2 & 5.5  & $2.8\times10^8$ & DE & Cubic & Adaptive \\
    \hline
    3HDCV &  3 & 1.5 &  $4.3\times10^6$ & DE & Cubic & Volume \\
    3MDCV &  3 & 3.0  & $3.5\times10^7$ & DE & Cubic & Volume \\
    3MDCA &  3 & 3.0  & $3.5\times10^7$ & DE & Cubic & Adaptive \\
    3MDQV &  3 & 3.0  & $3.5\times10^7$ & DE & Quintic & Volume \\
    3MPQV &  3 & 3.0  & $3.5\times10^7$ & PE & Quintic & Volume \\
    3LDCA$^\dagger$ &  3 & 5.5  & $2.8\times10^8$ & DE & Cubic & Adaptive \\
    \hline
  \end{tabular}
  \caption{Simulation matrix. D: density-entropy formulation; P:
    pressure-entropy formulation; V: radius of feedback region fixed to 
    0.5 physical $h^{-1} \mathrm{Kpc}$; A: feedback region 
    is set to 64 nearest neighbour (cubic spline) or 224 nearest neighbour 
    (quintic spline); H, M, L
      stands for the resolution (high, medium and low, see text); 
     Q: quintic spline kernel; C:
        cubic spline kernel.}
    \flushleft $\dagger$ the LDCA simulations use the same accretion and feedback model as
MassiveBlack.
  \label{tab:simulations}
\end{table*}

\subsection{Blackhole Accretion Model}
The blackhole accretion model was introduced by \cite{2005MNRAS.361..776S} and
\cite{2005Natur.433..604D}. Here we summarize some relevant features.
The blackholes are seeded in the simulations 
on-the-fly in newly formed FOF halos with $M_\mathrm{Halo} 
\le 5\times 10^{10} h^{-1}M_\odot
$. The seed mass is $M_\mathrm{bh,seed} = 5\times 10^{5} h^{-1}M_\odot$.
The accretion rate of a blackhole is calculated from
\[
  \frac{dM}{dt} = 4 \pi \cdot \mathrm{min}
  \begin{cases}
    A \rho(x)c_\mathrm{eff}(x)^{-3} (GM)^2 & \mbox{Bondi-Scaling } \\
    B m_\mathrm{p} (\eta \sigma_{T} c)^{-1} (GM) & \mbox{Eddington Limit} 
    \end{cases},
\]
where the local gas density $\rho(x)$ and local effective sound speed
$c_\mathrm{eff}(x)$ are evaluated from the nearest neighbour SPH 
estimate at the blackhole position $x$. 
$\sigma_{T}$ is the Thompson cross section, $G$ the
gravitational constant and $m_\mathrm{p}$ the proton mass. 
We have absorbed other constants in $A$ and $B$. 
The light to mass ratio $\eta$ is taken as $10\%$.
We also assume that $5\%$ of the light emitted becomes thermal feedback 
energy.

Two blackholes are merged when two nearby
(within the SPH resolution) blackholes
satisfy the condition  $v_\mathrm{rel} < {1}/{2}c_\mathrm{sound}$,
where $v_\mathrm{rel}$ is their pairwise relative velocity and
$c_\mathrm{sound}$ is the sound speed of the surrounding gas.

\subsection{Merger and accretion history}
\begin{figure*}
  \includegraphics[width=5in]{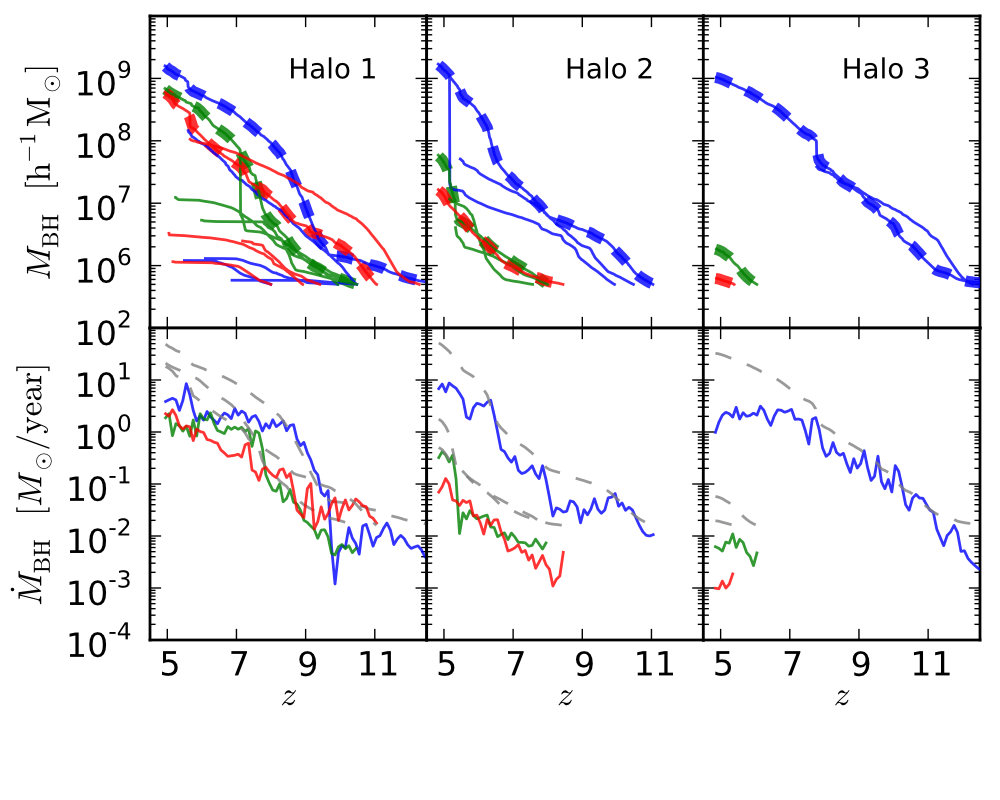}
  \caption{The merger tree of three most massive blackholes in each simulated
  halo. The data has been taken from the simulations with the highest
resolution (HDCV). From left to right we show halos 1, 2 and 3. The thick lines
show the most massive progenitors in each case; blackholes in the same merger
tree are shown with the same color. The lower panel shows the
accretion rate of the most massive progenitors; grey lines indicate the Eddington
rate.}
  \label{fig:bh-mergers}
\end{figure*}

When examining the accretion history of blackholes
in each zoom region in Sections 3.3 onward we will restrict ourselves
to the  most-massive progenitor of the most-massive blackhole
at $z=6$. We find
that halo 2 and halo 3 contain one central
blackhole each at the end of the simulation, while halo 1 contains three major
blackholes of similar mass. This can be
seen in Figure \ref{fig:bh-mergers} where we show the
mass evolution of the 3 most massive (final redshift)
blackholes and their progenitors in each halo. 
The difference in the environments
of the blackholes seen in Figure \ref{fig:halo-env}
shows up strikingly here too, with for example halo 3 hosting only one
early merger events between blackholes at close to $z=8$.

We note that where there are differences in the precise ranking of
blackholes for different versions of a simulation
we define as the central blackhole the
most massive blackhole in the highest resolution simulation.

In the following
subsections we explore the effect of varying the accretion and
feedback models 
on the growth of the central blackhole.  

\subsection{Deposition of AGN feedback energy}
\label{sec:region}
\begin{figure*}
  \includegraphics[width=5in]{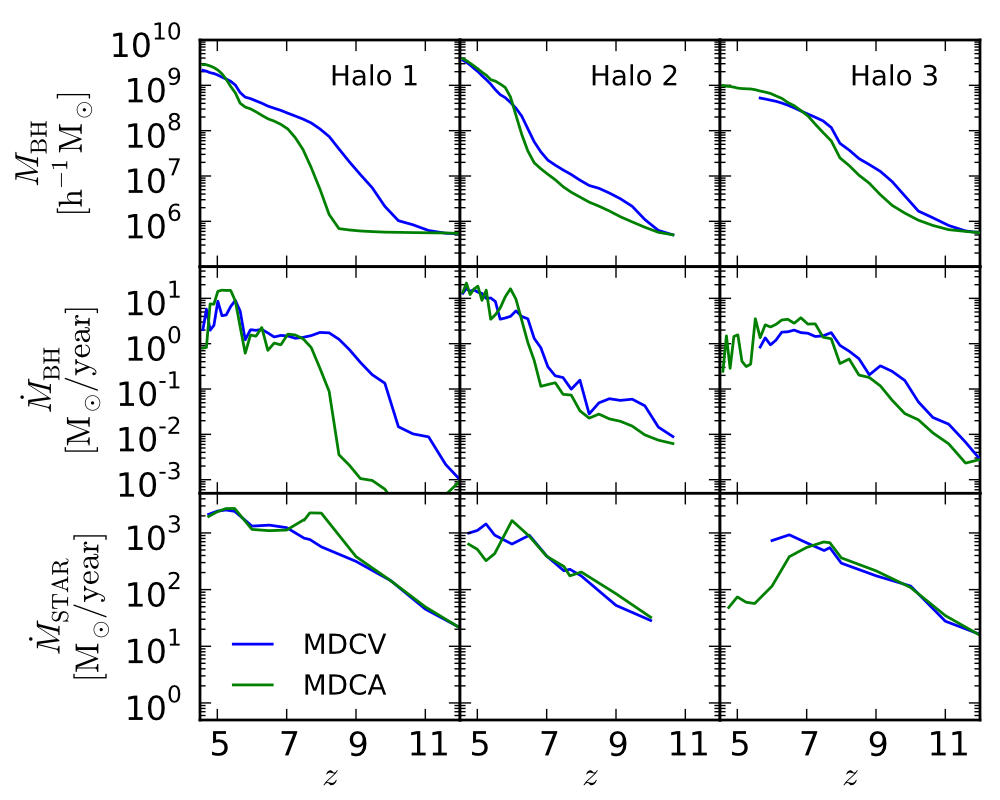}
  \caption{The dependence of blackhole accretion history on feedback energy
    deposition prescription. From left to right we plot halo 1, 2, 3. 
    The blue curves show results for the fixed-volume (finite physical radius
    of $0.5 h^{-1}\unit{Kpc}$) feedback kernel. The green curves are for the
  fixed-mass kernel, which uses the 64 nearest neighbours.  We also show the
  star formation rate of the halo in the bottom panels. Refer to the text in Section
  \ref{sec:region}.}
  \label{fig:radius}
\end{figure*}
The blackholes deposit feedback energy into nearby gas environment. In our
simulations, we model this process by depositing thermal feedback energy to the
neighbouring gas particles, weighted by their mass.
The exact details
of the physical mechanism by which this 
occurs in real galaxies is likely dictated by radiative transfer through the
medium surrounding the blackhole \citep{2009ApJ...701L.133A,
  2007ApJ...665.1038C,2006ApJS..165..188H, 2012ApJ...754...34J,2012MNRAS.427.2734N,
2009ApJ...696L.146M,2011ApJ...739....2P, 2012ApJ...747....9P}.
As the length and
mass scales over which this
occurs are not certain, it is important to test how  
our sub-grid prescription for depositing this feedback energy
affects growth of the blackholes.

In the original MB simulation the energy deposition was done using the SPH
smoothing kernel of the blackhole particle to distribute energy to the 64 nearest
neighbors. We refer to the MB feedback method ``adaptive'', as the size of the
feedback region is directly proportional to the mean separation of SPH particles
close to the blackhole (a ratio of $\eta=1.26$). In order to preserved the ratio, 
we use 224 neighbours in the quintic spline kernel simulations.

An alternative to the ``adaptive'' model is to fix the proper volume of the gas
receiving the feedback energy.  For the fixed-volume model, we use a region
corresponding to a spline smoothing kernel with proper radius $h =
0.5\,h^{-1}\unit{Kpc}$. The total mass of gas receiving the feedback energy is
therefore proportional to the gas density around the blackhole.  
Note that we do not use the fixed-volume model at low
resolution as the feedback region is smaller than the gravitational smoothing
when even at very high redshift ($z<10$).

In Figure \ref{fig:radius}, we compare the results of simulations with the
adaptive model for distributing feedback energy and those with the fixed-volume
model at the medium resolution (MDCA and MDCV simulations in Table
\ref{tab:simulations}). 
We can see that the blackhole
mass, accretion and the halo star formation in all three halos are very similar
for both feedback models (within the same order of magnitude). 
The only exception is in halo 1, where there is some difference in the early
growth: the accretion rate in the adaptive model remains 2 orders of magnitude lower
than the fixed-volume model until redshift $z=8.5$. The difference in accretion
rate disappears after $z<8$, and eventually the blackhole mass between two
models become very similar. The star formation in the halo also experiences a
burst during the same period, indicating an accumulation of cold dense gas in the
halo before the blackhole accretion kicks in. After the accretion is started, 
the blackhole mass quickly picks up. We therefore regard the similarity in
the late time blackhole masses and accretion rates as more definitive.
The choice of feedback region therefore does not appear to significantly alter
the accretion history.

\subsection{Resolution}
\label{sec:resolution}
\begin{figure*}
  \includegraphics[width=5in]{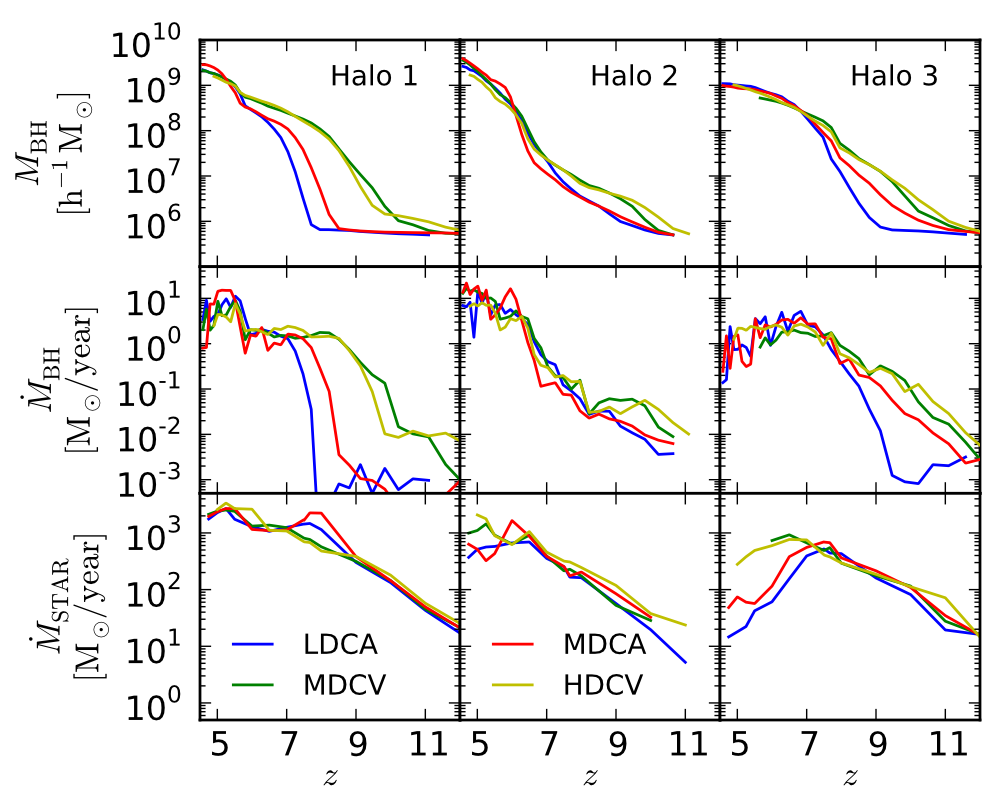}
  \caption{Resolution dependence of the accretion history of the most massive
    blackholes in our 3 selected target halos. From left to right we plot Halo
    1, Halo 2, and Halo 3. The blue (LDCA) curves are the lowest resolution, 
    green and red (MDCV, MDCA) medium resolution and yellow (HDCV) 
    is high resolution (see Section \ref{sec:resolution}).} 
    \label{fig:resolution}
\end{figure*}
The MB simulation has a gravitational force softening
length of $\varepsilon = 5.5\,h^{-1}\unit{Kpc}$. 
The re-simulations are conducted with three different gravitational force
softening lengths (in comoving coordinates):
$5.5\,h^{-1}\unit{Kpc}$ (low resolution),
$3.0\,h^{-1}\unit{Kpc}$ (med resolution), and
$1.5\,h^{-1}\unit{Kpc}$ (high resolution). Our highest resolution simulation
therefore has a proper resolution of $300$pc at $z=6$.

We plot the accretion history for the 3 different
blackholes at different resolutions in Figure \ref{fig:resolution}, showing
blackhole mass and accretion rate and the star formation rate in the halo. 
We find that the resolution does not appear to be affecting the blackhole
growth and the star formation in the halo. 
The accretion history and star formation appear to have reached convergence with
the medium resolution simulations(MDCV and MDCA), which shows no significant
difference with the high resolution simulations (HDCV) that has 8 times better
mass resolution.

The major difference again lies in the adaptive feedback model. 
We see that the initial growth ($z > 8$) of the adaptive model shows
stronger resolution dependence than the fixed-volume model in both halo 1 and
halo 3: the lower resolution
simulation (LDCA) produces slower initial growth at $z > 8$ than the higher
resolution (MDCA). The difference in halo 3 is less drastic than the difference
in halo 1 and is not associated with a burst in star formation rate. We believe
this is because the halo is in a relatively isolated region, and the cold gas
supply is lower than halo 1.

We point out that eventually ($z < 6$) the lower resolution simulation
(LDCA) does also reach the same final blackhole mass as the medium resolution
simulation (MDCA), with the dependence on resolution disappearing.
This is because at $z<6$ the blackhole is sufficiently massive that the 
accretion becomes regulated by feedback, regardless of how the feedback is
implemented.

\subsection{SPH formulation: Pressure-Entropy}
\begin{figure*}
  \includegraphics[width=5in]{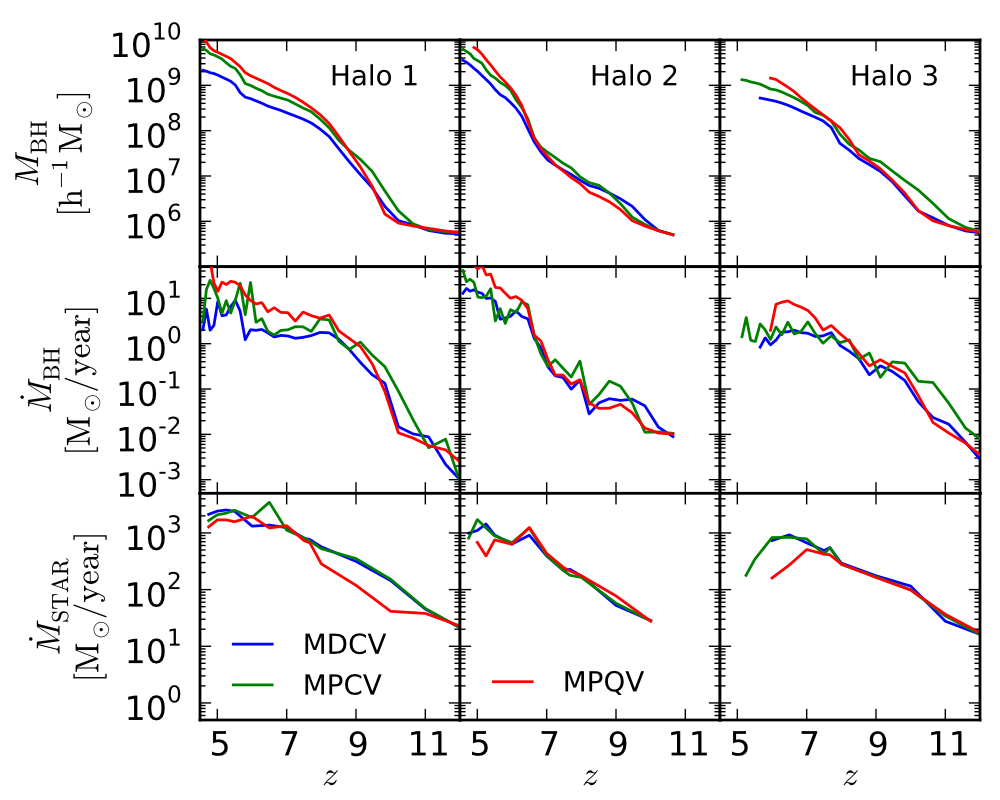}
  \caption{The effect of changing SPH formulation on the accretion history of
    the blackholes.  From left to right we show halo 1, halo 2, and halo 3. The
    red (MDCV) line is the density-entropy formulation with a cubic spline
    smoothing kernel; the green (MPCV) line is the pressure-entropy formulation
  with a  cubic spline smoothing kernel and the blue (MPQV) line is the
pressure-entropy formulation with quintic spline smoothing kernel.}
  \label{fig:pesph}
\end{figure*}

Recent developments in SPH include the introduction of the
so-called pressure-entropy formulation \citep{2013MNRAS.428.2840H,
2010MNRAS.405.1513R}, and the quintic smoothing kernel
\citep{2008JCoPh.22710040P}. These new developments aimed to address several
difficulties noticed in prior formulations of SPH, namely:  
an unphysical surface tension across surface boundaries that forbids particle
exchange at a density discontinuity, 
the fluctuation in density estimation due to a small number of nearest
neighbours (32), and nearby particles bonding into pairs causing a loss of
resolution in high density regions \citep[see e.g.][]{2007MNRAS.380..963A,
1996PASA...13...97M,1981A&A....97..373S,
2000JCoPh.159..290M,2008MNRAS.387..427W}.

We have run
our zoomed simulations with pressure-entropy SPH
and the traditional density-entropy SPH formulation.
The blackhole accretion histories and the halo star formation rates are
shown in Figure \ref{fig:pesph}. We can see that 
going from density-entropy to pressure-entropy formulations appears to
consistently result in faster blackhole accretion, while the star formation is
not affected.  Comparing the results for quintic spline and cubic kernels 
(both are plotted) we can 
see that some of the contribution is due to
the switch to a quintic spline kernel. The quintic spline kernel 
samples more particles (112)
and reduces
the noise in the quantities used the blackhole accretion model. The blackholes
are up to a factor of 5 times more massive at the final time
in the pressure-entropy case compared to density entropy case. 
We note that there is a free parameter in the blackhole model, namely
the feedback efficiency (set to 10 $\%$ here). This parameter
was set \citep{2005Natur.433..604D} to reproduce blackhole masses observed
in the local Universe and so with pressure-entropy SPH a recalibration
of this parameter could be employed.

\section{Feeding Blackholes via Cold Flows} 
\begin{figure*}
  \includegraphics[width=2in]{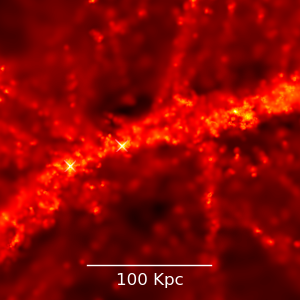}%
  \includegraphics[width=2in]{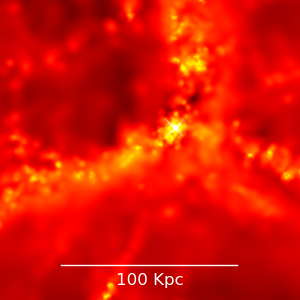}%
  \includegraphics[width=2in]{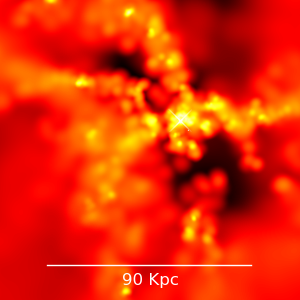}
  \includegraphics[width=2in]{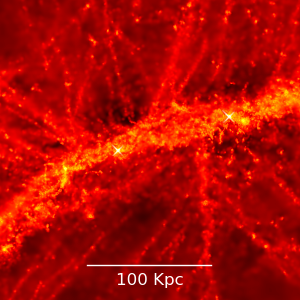}%
  \includegraphics[width=2in]{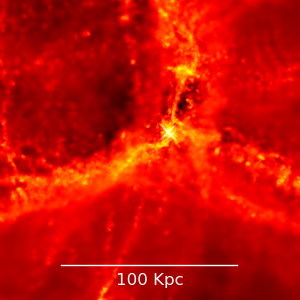}%
  \includegraphics[width=2in]{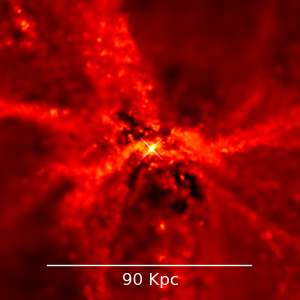}
  \includegraphics[width=2in]{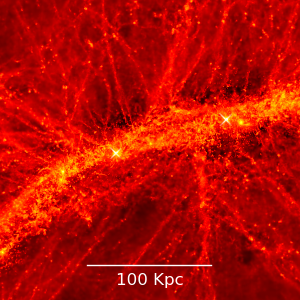}%
  \includegraphics[width=2in]{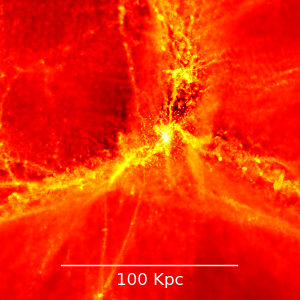}%
  \includegraphics[width=2in]{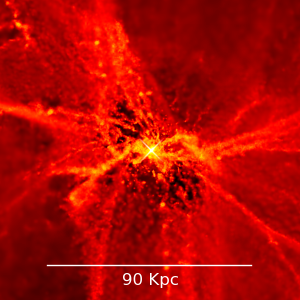}
  \includegraphics[width=2in]{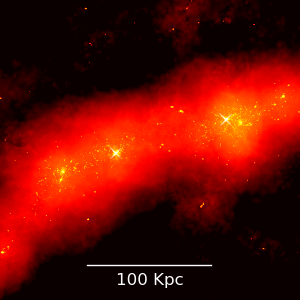}%
  \includegraphics[width=2in]{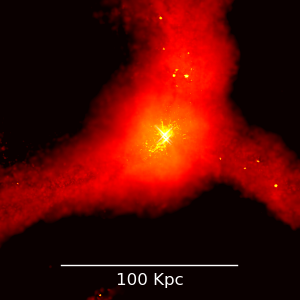}%
  \includegraphics[width=2in]{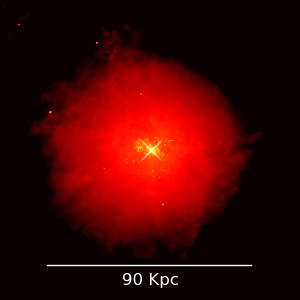}
  \caption{Top 3 rows:
Cold gas filaments surrounding blackholes in the simulations,
run at different resolution. The color represents the
    projected density (in a cube) 
of cold  ($T < 10^5\unit{K}$) gas at $z=6.5$. 
    The columns show Halos 1, 2, and 3, from left to right.
    The first 3 rows  are, from top to bottom, low resolution, medium
 resolution and high resolution. 
  The bottom row shows the 
density distribution of the hot gas in the high resolution simulation.}
  \label{fig:cold-flow-visual}
\end{figure*}

We now investigate the fuel supply onto blackholes and 
study the thermal history of high redshift gas that participates in
blackhole accretion. We focus on the question of whether the gas that fuels
the black hole is different from the typical gas in the halo.
We want to further test the indication that at these high redshifts ($z\sim6$)
fast black hole growth can be achieved via cold flows
\citep{2012ApJ...745L..29D}. Further we aim to investigate how black hole
feedback may play a role in disrupting the cold streams close to the black
hole. 

The virial temperature of the hosting halo of $10^{12} \unit{M_\odot}$ is close
to $10^6 \unit{K}$, while in the simulation the cold IGM temperature is close to
$10^4 \unit{K}$. We set the split between hot and cold gas at $10^5\unit{K}$. 
In Figure \ref{fig:cold-flow-visual}, we compare the \ed{morphology} of 
the cold gas ($T < 10^5\,
\unit{K}$) and that of the hot gas ($T > 10^5\, \unit{K}$)
around the blackholes in the three halos. We first notice that
the cold gas forms dense compact filaments that are unlikely due to resolution
effects: as we increase resolution, the cold gas filaments are identical in
morphology except have increased sharpness.  Secondly, we notice that the hot
gas is distributed in a diffuse manner, forming thicker structures containing
large blobs that extend to the entire halo.

The striking morphological difference between hot and cold gas motivates us to
consider two species of gas particles:
\begin{enumerate}
  \item the first gas species we call ``HALO'' 
  particles\footnote{Not to be confused with the dark matter particles in the
halo.}
  and are randomly selected gas particles from the halo, weighted
  by volume; because hot gas particles occupy most of the volume
  of the halo, HALO particles are most likely the hot environment gas in
  the halo shown in the bottom panel of Figure \ref{fig:cold-flow-visual}.
  \item the second gas species are the ``BH'' 
   particles\footnote{Not to be
    confused with particles that represents the blackholes in the
simulation.}, and are the
  nearest neighbour particles contributing to the evaluation of gas properties at
  the blackhole. The BH particles are the gas particles that are
  actively participating in accretion onto the blackhole.
\end{enumerate}

After a gas particle (if it has not been converted to a star particle) enters
the halo, depending on its cooling efficiency, the particle will either be
heated to the virial temperature of the halo, losing its initial inflow motion,
or remain cold, arriving to the center of the halo to participate in star
formation and blackhole accretion. In this picture, the history of a gas
particle can be quantified by two characteristic times:
\begin{enumerate}
  \item the entering time $t_\mathrm{enter}$,  the time since
the Big Bang at which a gas particle first enters
    the virial radius; 
  \item the heating time $t_\mathrm{heated}$, the time since the
Big Bang at which a gas particle is first heated above the
  virial temperature of the gas.
\end{enumerate}
We also calculate the correlation coefficient (CC) of the two characteristic
times  
\[
    \mathsf{CC} = \rho_{X,Y}={\mathrm{cov}(X,Y) \over \sigma_X \sigma_Y}
={E[(X-\mu_X)(Y-\mu_Y)] \over \sigma_X\sigma_Y},
\]
where $X$ and $Y$ represent $t_\mathrm{enter}$ and $t_\mathrm{heated}$, 
$\sigma$ and $\mu$ the standard-deviation and the mean of the data,
and $E$ the mean operator.

CC measures the correlation between the time a gas particle enters the halo and
the time a particle is heated to the virial temperature of the halo. If a
particle is heated by the virial heating from the halo, we expect CC to be
large, for particles entering the halo early shall be heated early, while those
entering late shall be heated late. On the other hand, if a particle is heated by
the central blackhole, we expected CC to be small, for these particles are
heated only after they arrive to the blackhole, no matter when they enter the
halo. As we will show later, these expectations agree with our simulations. 

\begin{table*}
\begin{tabular}{ccccccc}
\hline
\hline
Halo     & \multicolumn{2}{c}{1} &\multicolumn{2}{c}{2} &\multicolumn{2}{c}{3} \\
Species  &  BH & HALO &   BH & HALO & BH & HALO \\
\hline 
EA (high redshift)  & 0.069 & 0.574 & -0.075 & 0.778 & 0.131 & 0.337 \\
RA ($z = 5.5$)  & 0.389 &  0.893 & 0.356 & 0.745 & 0.557 & 0.827 \\
\hline
\hline
\end{tabular}
\caption{Correlation coefficient (CC) between $t_\mathrm{heated}$ and
$t_\mathrm{enter}$. Also see Figure \ref{fig:two-phases}.}
\label{tab:two-phases}
\end{table*}

We calculate $t_\mathrm{enter}$, $t_\mathrm{heated}$, and CC for BH and HALO
particles at two epochs from the high resolution simulations (HDCV): (1) the
first epoch is when the blackhole is undergoing Eddington accretion (EA phase,
$z=9\,,6.5\,,\text{and}\,7.5$ for each halo respectively), and (2) the second
epoch is when the blackhole accretion has been regulated (RA phase, $z=5.5$ for
all three halos). 

\begin{figure*}
  \includegraphics[width=\textwidth]{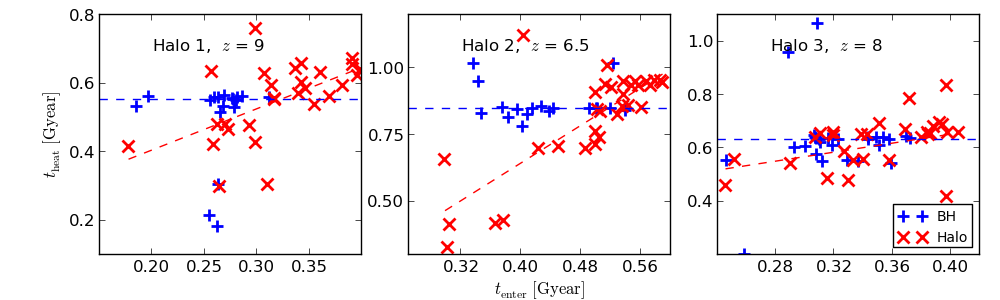}
  \includegraphics[width=\textwidth]{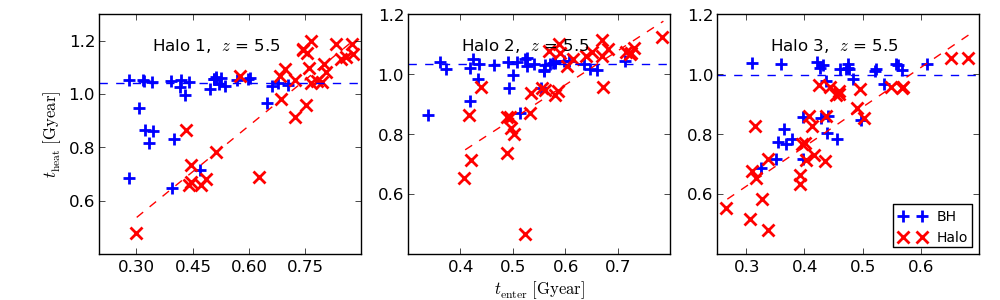}
  \caption{Time since the Big Bang that a particle
enters a halo ($t_{\rm enter}$)
plotted against time since the Big Big for a particle
to heat up to the virial temperature ($t_{\rm heat}$)
HALO particles and BH particles (see text) are shown in red and blue respectively.
    The panels show halos 1, 2 and 3 from left to right. The upper panels are at
    the EA phase when the blackholes are undergoing Eddington accretion. The
    lower panels are in the RA phase when the blackhole growth is regulated by
    feedback.  The red horizontal dashed line indicates the time when most of 
    the BH particles are heated, acting as a visual guide to the timescale of 
    blackhole heating. The blue dashed line is a linear fit to the HALO
    particles, visually guide the halo heating.}
  \label{fig:two-phases}
\end{figure*}

The results are shown in Figure \ref{fig:two-phases} and Table
\ref{tab:two-phases}. We first look in the EA phase. During the EA phase, the
correlation coefficient (CC) between the heating time and entering time of the
BH particles is much smaller than that of the HALO particles as seen in the
first row in Table \ref{tab:two-phases}. The difference can also be seen in
Figure \ref{fig:two-phases}, where the BH particles are clustering around the
same $t_\mathrm{heated}$ that corresponds to the time the particle becomes the
nearest neighbour of the blackhole, while the HALO particles follows a line with
a positive slope. The lower CC in BH particles is consistent with heating by
blackhole feedback: the BH particles remain cold after they enter the halo,
and are then heated up shortly after they arrive at the blackhole.  The larger
CC with HALO particles in EA phase is consistent with the virial heating
picture: the hot gas halo is formed by virial heating from the halo (which may
indirectly contain the feedback energy from the central blackhole and the
feedback energy from satellites); if a HALO particle enters the halo early, it
is also heated early. 

This difference suggests that the gas that participates the
blackhole accretion in an Eddington accretion phase is unlikely to be
from the hot halo environment. \ed{The gas must have come directly from outside
of the halo; and its thermal history is consistent with that of the cold flows. }

We then move on to the RA phase, during which blackhole accretion has
been regulated.  We notice that the CC of the BH particles have significantly
increased. In the lower panels of Figure \ref{fig:two-phases}, we find this
increase is due to a larger population of the BH particles that behaves
similarly to the HALO particles.  The existence of a HALO-like
population within the BH particles shows that the cold flows (though
they still exist) are being somewhat disrupted. Because the HALO-like particles
are heated before they arrive to the blackhole, the temperature around
the blackhole rises, and the blackhole accretion is regulated.

\begin{figure*}
  \includegraphics[width=\textwidth]{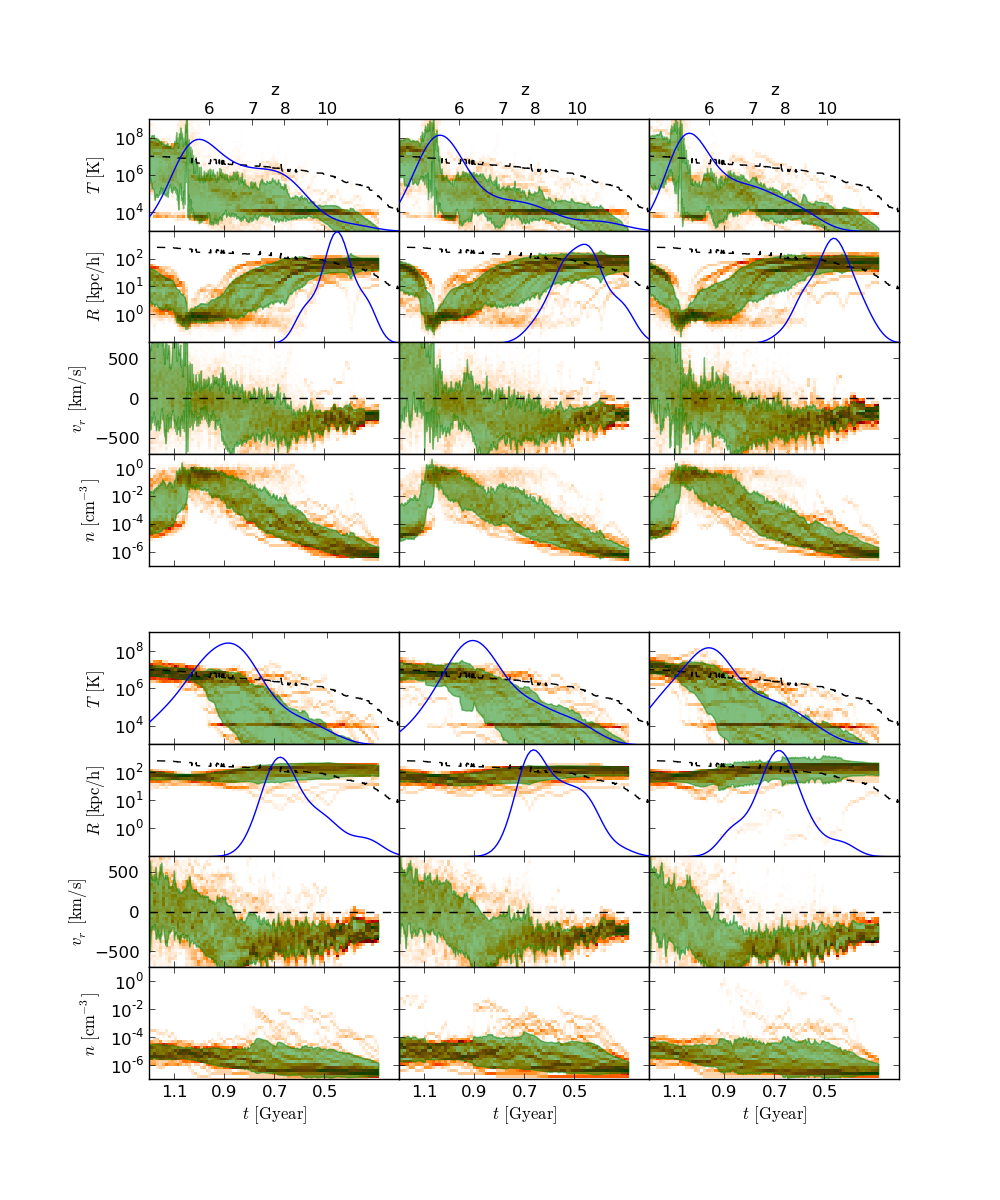} %
  \caption{The properties of HALO and BH particles selected from three snapshots
          close to $z=5.5$, as function of time. 
  Top row   : BH particles (nearest neighbours of the blackhole). 
  Bottom row: HALO particles (randomly picked from inside the halo).
  The sub-panels with in a plot are: 
    temperature ($T$), distance to blackhole ($r$), radial velocity ($v_r$) 
     and density ($n$).
  The 2-d histogram visualizes the trajectories of particles. (See text) 
  The 1-$\sigma$ contour of the distribution of the traced properties 
  is shown in green.
  The black dashed lines mark the virial temperature, virial radius, and 
  zero radial velocity. 
  The blue solid lines on $T$ and $r$ panels show the probability 
   distribution of the crossing time for temperature ($t_\mathrm{heated}$) and
  radius ($t_\mathrm{enter}$) (see text). 
  }
  \label{fig:histories}
\end{figure*}

\begin{figure*}
  \includegraphics[width=\textwidth]{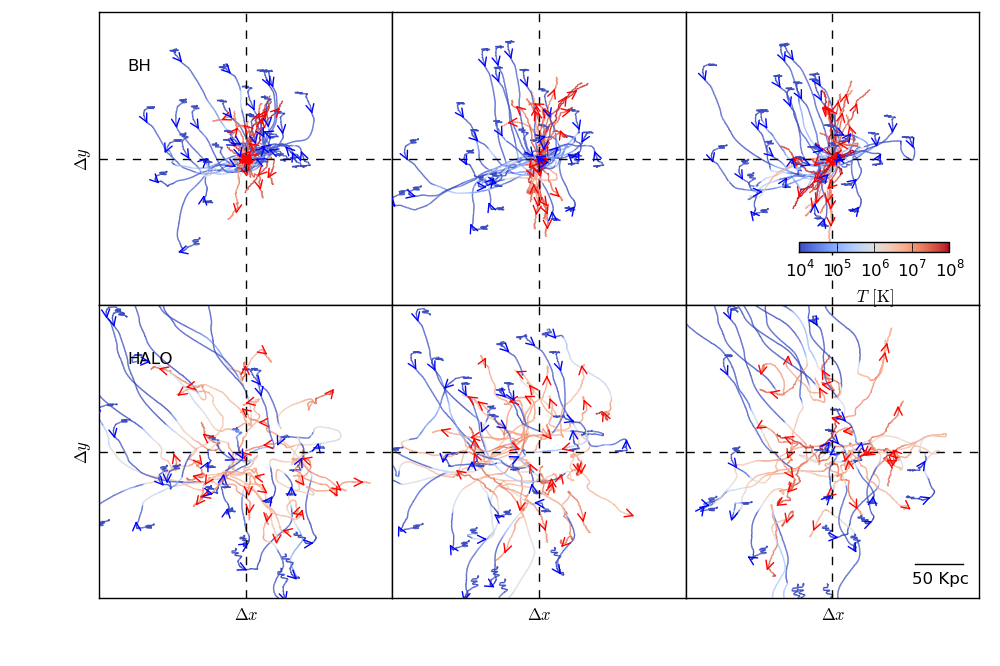}
  \caption{The trajectories of BH and HALO gas particles, projected into
the $x-y$ plane, during the RA phase. 
    Top panels: BH particles that participate in accretion onto the blackhole;  
    Bottom panels: HALO particles.  
    The trajectories are color-coded by temperature:
    red is $10^8\unit{K}$, blue is $10^3\unit{K}$. 
   The wedges shows the direction of motion. 
  }
  \label{fig:motion}
\end{figure*}

We investigate further the motion of BH particles and HALO particles in the RA
phase with Figure \ref{fig:histories} and Figure \ref{fig:motion}.
In Figure \ref{fig:histories}, we plot the history of properties of BH and HALO
particles in simulation 1MDCV, where we saved one snapshot every $3\,
\unit{Myear}$ of simulation time. In this case we have traced 32 particles of
each population originating from three snapshots close to $z=5.5$ (three columns
are three snapshots; two rows are BH and HALO from top to bottom, note that the
time increases from right to left). The
considered properties are:
\begin{description}
\item[$T$:] the temperature of the particle; we compare the temperature with
the virial temperature of the halo.
\item[$r$:] the distance between the particle and the blackhole (proper); we compare
the distance with the virial radius of the halo.
\item[$v_r$:] the radial velocity of the particle; negative value ($v_r < 0$) is
infalling and positive ($v_r > 0$) is out-flowing.
\item[$n$:] the density of the particle.
\end{description}

\begin{figure*}
  \includegraphics[width=\textwidth]{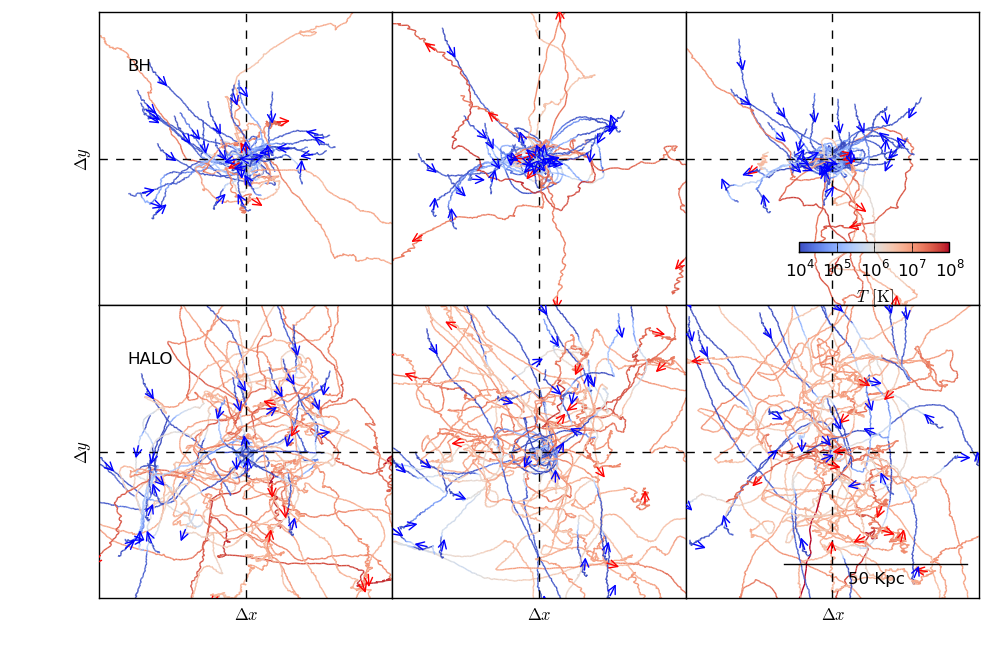}
  \caption{The trajectories of BH and HALO gas particles, projected into
the $x-y$ plane, during the EA phase. 
    Top panel: BH particles that participate in accretion onto the blackhole;  
    Bottom panel: HALO particles.  
    The trajectories are color-coded by temperature:
    red is $10^8\unit{K}$, blue is $10^3\unit{K}$. 
    The wedges shows the direction of motion. 
  }
  \label{fig:motion-EA}
\end{figure*}

We visualize the history of each property with a 2-d histogram binned by the
quantity and time in Figure \ref{fig:histories}. The histogram demonstrates the
behavior of the entire population as a function of time: 
the particles with similar property at a given time will contribute to the same
bin, resulting a darker pixel. For example, the dark horizontal line at
$T=10^4\unit{K}$ in the top
panel $T$ plots of Figure \ref{fig:histories} indicates a large fraction of gas
remains cold for a very long time.
We also mark the probability distribution of 
$t_\mathrm{heated}$ and $t_\mathrm{enter}$ in the figure. We are more interested
in the behavior of the gas between $t_\mathrm{enter}$ and $t_\mathrm{heated}$.

In the three top panels for BH particles, 
we see that most of the BH particles have significantly increased their density
while remaining cold for as long as $0.5\,\unit{Gyear}$ after they enter the
halo, as suggested by the dark lines in $T$, $v_r$ and $n$. We do notice
that for some of the particles, the infalling motion starts to become
disrupted much sooner ($0.2\,\unit{Gyear}$ after entering), while their
temperature rise to the virial temperature. As we will see after analyzing the
HALO particles, this early heating population within BH gas are very similar to
the HALO particles: they correspond to the HALO-like population we identified in
the lower panels of
Figure \ref{fig:two-phases}, which are responsible for the increased correlation
between $t_\mathrm{enter}$ and $t_\mathrm{heated}$. The
population indicates that the cold flows in a regulated growth phase are being
disrupted, and can no longer provide the Eddington accretion of blackholes.
All of the BH particles are heated after they get close to the blackhole ($r <
1\,\unit{Kpc}$), but the temperature reaches up to $10^8\,\unit{K}$, much higher
than the virial temperature of the halo ($10^7\, \unit{K}$) at this redshift. 
We also note that
some of the heated particles can receive an out-flowing velocity up to
$1000\,\unit{km/s}$, and can reach a substantial distance away from the
blackhole.

In the three low panels for HALO particles, we see that the heating happens much
sooner (within $0.2\,\unit{Gyear}$) after they enter the halo. The infalling
motion is disrupted as the particles are heated, and most of the particles
remains at a low density of $10^{-6} \unit{cm^{-3}}$. After being heated, the
temperature of HALO particles tightly tracks the virial temperature of the halo,
as the halo grows. 

We proceed to visualize the motion of the in-flowing gas with Figure
\ref{fig:motion} for the particles selected from the three snapshots near
$z=5.5$. In Figure \ref{fig:motion}, we show the trajectories of 
the BH and HALO particles projected in the $x-y$ plane, colored by their
temperature. The BH particles, being initially cold ($T \sim 10^4\unit{K}$),
flow into the blackhole (located at the center of the plots), apparently
following several preferred directions. The majority of the BH particles remain
cold until they arrive to the blackhole. However, we can also see some of the
particles are heated to $T>10^6\unit{K}$ before arriving the the blackhole; 
these particles correspond to the HALO-like particles.
Afterwards, the particles are ejected towards directions that are unaligned to the
directions they come in from. The ejected gas can reach  more than $100
\,\unit{Kpc}$ away from the blackhole.
The HALO particles show a very different pattern. Although the HALO particles
are also initially cold ($T \sim 10^4\unit{K}$), the gas is quickly heated 
to the virial temperature of the halo ($T \sim 10^6 \,\unit{K}$). Afterwards,
the motion of particles visually resembles a random-walk within the virial radius
of the halo (close to $100 \,\unit{Kpc}$).

The fate of BH particles in the EA phase is drastically different. As shown in 
the top panels of Figure \ref{fig:motion-EA}, we see that instead of being
ejected away, most of the BH particles remain cold after arriving at the
blackhole and they are eventually consumed by blackhole accretion and star
formation at the very center of the halo. The fate of HALO particles in the EA
phase is rather similar to that in the RA phase: particles are heated to the
virial tempearture as they fall into the halo.

\section{Conclusion}
We have studied the fueling and gas supply onto high-redshift supermassive
blackholes in high resolution re-simulations of halos selected from the 
MassiveBlack hydrodynamic cosmological simulation (covering a volume close to
1 Gpc$^3$). Using MassiveBlack, Di Matteo et al. (2012) showed that steady high
density cold gas flows were able to produce the high gas densities that can
lead to sustained critical (Eddington) accretion rates and hence rapid growth
commensurate with the existence of $~10^9 M_{\sun} $ blackholes as early as
$z \sim 7$. We have tested this scenario further for a subsample of three of
the halos hosting $10^9$ solar mass blackholes at $z =
6.0$ in MassiveBlack.  We have used zoom-in techniques to investigate the nature
of the gas inflows at scales that could not be resolved from the large volume
MassiveBlack simulation.

We have shown that for the three re-simulated halos the growth history of their
central super massive blackholes is consistent with a sustained Eddington
limited fueling provided by cold streams of gas that penetrate all the way to
the innermost region of the galaxies undisrupted (i.e. consistent with our
earlier picture). Our conclusions remain unchanged even though the numerical
scheme we use is varied in several significant aspects:
\begin{itemize}
  \item a change in resolution does not change the fuelling history; 
  \item AGN feedback
 energy is spread over different size regions of constant mass or volume, 
\item an improved formulation of SPH (pressure-entropy and quintic kernel) is
  used (in this case, the growth of the blackholes is systematically faster).
\end{itemize}

Our results are consistent with
the recent work by \cite{2013arXiv1307.5854C} that shows that out of 18 halos 
selected in the Millennium simulation (dark matter only)
only six halos found in large scale over dense regions are able to 
experience critical growth commensurate with $10^9 M_\odot$ by $z=6$.  
Recently, \cite{2013arXiv1307.0856B} also studied the growth history of black
holes in high redshift halos. Their zoomed regions were extracted from an
initial cosmological box of only $50\, \mathrm{Mpc}$ per side, unlikely to
contain high enough overdense regions. The final blackhole
masses in their work are two orders magnitude smaller, consistent with what
is expected from more moderately over-dense regions, as studied by
\cite{2013arXiv1307.5854C}. 

We have investigated the nature of the gas fueling the central
blackholes by computing the correlation between the time required for 
gas particles (in the halo) to reach the virial temperature and the time since 
they first enter the halo. We also investigated the correlation of gas participating
in accretion and the correlation of the hot gas in the halo environment. 
We have found that while gas that fuels the blackhole preserves its
low temperature until heated by the feedback from the blackhole (low correlation), 
the gas ending up in the halo environment typically heats quickly upon entering
the halo (high correlation).  
With this method, we have shown that during the Eddington growth
phase, the accretion gas appears to have directly arrived to the blackhole 
through cold flows without disruption from the hot environment; 
while during the regulated growth phases ($z < 6$) a larger population of the
accretion gas behave very similar to the hot environment, indicating that the
cold flows are becoming disrupted by the feedback.

After being heated by the blackhole the gas particles are ejected with a high
velocity up to $1000 \unit{km/s}$ from the blackhole.  Note also that several
\ed{authors \citep[see, e.g.][]{2013MNRAS.428.2885D,2011MNRAS.412.1965M}
reported that the strong outflows during the regulated phase at high redshift
can reduce the baryon fraction in the halo up to 30\%, increasing the entropy of
halos at lower redshift. } In our halos, the disruption appears to be not as
strong, as most of the gas surrounding the blackhole is cold even after the
accretion is regulated. We do note that in our smallest halo (halo 3) have
the largest HALO-like population, indicating that the feedback from super
massive blackholes may be capable of blowing the cold flows away in less massive 
halos such as those studied by these authors.

\section*{Acknowledgements}
The resimulations used in this work and the 
MassiveBlack simulation were run on the
Cray XT5 supercomputer Kraken at the National Institute for Computational
Sciences.  We acknowledge support from Moore foundation which enabled us to
perform the data analysis at the McWilliams Center of Cosmology at Carnegie
Mellon University.  This research has been funded by the National Science
Foundation (NSF) PetaApps programme, OCI-0749212 and by NSF AST-1009781.  The
visualizations were produced using GAEPSI (Feng et al. 2011).

\bibliographystyle{mn2e}
\bibliography{mybib}    
\end{document}